\documentstyle[sprocl]{article}

\def\be{\beta}
\def\ga{\gamma}
\def\de{\delta}
\def\ep{\epsilon}

\def\et{\eta}
\def\th{\theta}

\def\la{\lambda}

\def\ph{\phi}

\def\ch{\chi}

\def\om{\omega}
\def\Ga{\Gamma}
\def\De{\Delta}

\def\La{\Lambda}

\def\Ps{\Psi}
\def\Om{\Omega}
\def\cA{{\cal A}}

\def\cN{{\cal N}}

\def\fr#1#2{{{#1} \over {#2}}}
\def\frac#1#2{\textstyle{{{#1} \over {#2}}}}
\def\half{{\textstyle{1\over 2}}}

\def\prt{\partial}

\def\ket#1{|{#1}\rangle}
\def\bra#1{\langle{#1}|}

\def\lsim{\mathrel{\rlap{\lower4pt\hbox{\hskip1pt$\sim$}}
    \raise1pt\hbox{$<$}}}
\def\gsim{\mathrel{\rlap{\lower4pt\hbox{\hskip1pt$\sim$}}
    \raise1pt\hbox{$>$}}}

\def\Re{\hbox{Re}\,}
\def\Im{\hbox{Im}\,}

\def\X{\hat X}
\def\Y{\hat Y}
\def\Z{\hat Z}
\def\x{\hat x}
\def\y{\hat y}
\def\z{\hat z}

\newcommand{\bequ}{\begin{equation}}
\newcommand{\eequ}{\end{equation}}
\newcommand{\beq}{\begin{eqnarray}}
\newcommand{\eeq}{\end{eqnarray}}
\newcommand{\bea}{\begin{eqnarray}}
\newcommand{\eea}{\end{eqnarray}}
\newcommand{\rf}[1]{(\ref{#1})}

\def\etal{\it et al.\rm}

\def\ol#1{\overline{#1}}

\begin{document}

\title{CPT AND LORENTZ VIOLATION IN \\  
NEUTRAL-MESON OSCILLATIONS\footnote{IUHET 448, presented at 
the Second Meeting on CPT and Lorentz Symmetry,
Bloomington, Indiana, August 2001.}}

\author{V.\ ALAN KOSTELECK\'Y}

\address{Physics Department, Indiana University\\
Bloomington, IN  47405, U.S.A.}

\maketitle\abstracts{ 
The status of CPT tests with neutral mesons
is reviewed in the context of quantum field theory 
and the Lorentz- and CPT-violating standard-model extension.
}

\section{Introduction}

Local relativistic quantum field theories,
including the standard model of particle physics,
are known to be invariant under Lorentz and CPT transformations.
This symmetry is consistent with the results 
of numerous sensitive laboratory tests. 
Although no definitive violation has been discovered to date,
there are many reasons to undertake careful theoretical studies 
of possible mechanisms and descriptions 
of Lorentz and CPT 
violation.\cite{cpt98}
One basic motivation is that 
a comparative and quantitative interpretation 
of the numerous experimental tests
requires a comprehensive theoretical framework
within which violations are both allowed and internally 
consistent.\cite{kp,ck}
A more ambitious motivation is that 
suppressed Lorentz and CPT violation might arise 
from a fundamental theory at the Planck 
scale\cite{kps,chklo,bek}
but nonetheless be observable with existing technology
in experiments of exceptional sensitivity.

At the 1998 Bloomington conference on CPT and Lorentz 
symmetry,\cite{cpt98}
I discussed the possibility that Lorentz and CPT symmetry
might be broken by physical effects arising 
in a theory underlying the standard model,
including string 
theory.\cite{kps}
I also described the general standard-model extension
allowing Lorentz and CPT 
violation\cite{kp,ck}
and summarized some of the experiments 
that had already been performed to test it at that time.
In the intervening three years,
substantial advances have been made
on both the theoretical and experimental fronts,
many of which are discussed in other presentations
at this meeting.
In particular,
experimental tests of the standard-model extension now include
studies of neutral-meson 
oscillations,\cite{kexpt}$^{\rm -}$\cite{ckvki,kp}
comparative tests of QED in Penning 
traps,\cite{eexpt}
spectroscopy of hydrogen and 
antihydrogen,\cite{hexpt}
measurements of muon
properties,\cite{muexpt}
clock-comparison 
experiments,\cite{ccexpt}
tests with spin-polarized matter,\cite{eexpt2}
measurements of cosmological 
birefringence,\cite{photexpt,jk}
studies of neutrinos,\cite{bpww}
and observations of the baryon 
asymmetry.\cite{bckp}
These experiments measure coefficients for Lorentz and CPT violation
in the standard-model extension
and are probing the Planck scale.

This talk focuses on the theoretical issues 
involving tests of the standard-model extension
using neutral-meson oscillations.
Meson interferometry 
is a sensitive tool for both CPT and Lorentz violation.
Any indirect CPT violation in a neutral-meson system
can be parametrized with a complex quantity,
denoted in this talk by $\xi_P$,
where $P$ is one of the neutral mesons
$K$, $D$, $B_d$, $B_s$.
The talk outlines the formalism involving $\xi_P$,
describes the calculation of $\xi_P$ 
in the general Lorentz-violating standard-model extension,
and briefly considers some implications for experiment.
Reports on the latest experimental results
in the $K$, $D$, and $B_d$ systems
are being presented separately at this
conference.\cite{kexpt,dexpt,bexpt} 
The reader may also find of interest
some related recent analyses 
of possible classical analogues for CPT violation in neutral 
mesons,\cite{osc}
which fall outside the scope of this talk.

\section{Setup}

Any neutral-meson state is  
a linear combination of the Schr\"odinger wave functions 
for the meson $P^0$ and its antimeson $\overline{P^0}$.
If this state is viewed as a two-component object $\Ps(t)$,
its time evolution is controlled by 
a 2$\times$2 effective hamiltonian $\La$ 
according to the Schr\"odinger-type 
equation\cite{lw}
\beq
i\prt_t \Ps = \La \Ps.
\label{seq}
\eeq
Note that the effective hamiltonian 
is different for each neutral-meson system,
but for simplicity a single symbol is used here.

The eigenstates of $\La$ are the physical propagating states 
of the neutral-meson system,
denoted here as 
$\ket{P_a}$ and $\ket{P_b}$.
These states develop in time according to 
\beq
\ket{P_a(t)}=\exp (-i\la_at) \ket{P_a},\quad 
\ket{P_b(t)}=\exp (-i\la_bt) \ket{P_b},
\label{timevol}
\eeq
as usual.
The complex parameters $\la_a$, $\la_b$ in these equations
are the eigenvalues of $\La$,
and they are comprised of the physical masses $m_a$, $m_b$ 
and decay rates $\ga_a$, $\ga_b$ 
of the propagating particles:
\beq
\la_a \equiv m_a - \half i \ga_a, \quad 
\la_b \equiv m_b - \half i \ga_b.
\label{mga}
\eeq
For practical purposes,
it is convenient to work instead with the 
sum and difference of the eigenvalues,
defined as
\bea
\la &\equiv &\la_a + \la_b = m - \half i \ga,
\nonumber\\
\De \la &\equiv &\la_a - \la_b = - \De m - \half i \De \ga.
\label{ldl}
\eea
In these equations,
$m = m_a + m_b$, $\De m = m_b - m_a$,
$\ga = \ga_a + \ga_b$, and $\De \ga = \ga_a - \ga_b$.

Since the effective hamiltonian is a 2$\times$2 complex matrix,
it consists of eight independent real quantities for each meson system.
Four of these can be specified in terms of the masses and decay rates.
Three of the others determine the extent of indirect CP violation in the 
neutral-meson system.
If (and only if) the difference 
$\De\La \equiv \La_{11} - \La_{22}$ 
of diagonal elements of $\La$ is nonzero, 
then the meson system exhibits indirect CPT violation.
Also,
indirect T violation occurs 
if (and only if) the magnitude of the ratio 
$|\La_{21}/\La_{12}|$
of the off-diagonal components of $\La$ differs from 1. 
The effective hamiltonian thus contains
two real parameters for CPT violation
and one real parameter for T violation.
The remaining parameter of the eight in $\La$ 
can be taken as the relative phase between 
the off-diagonal components of $\La$.
It is physically irrelevant
because it can be freely changed 
by shifting the phases of the $P^0$ and $\overline{P^0}$ wave functions
by equal and opposite amounts.
Such shifts are allowed
because the wave functions are strong-interaction eigenstates.
If the $P^0$ wave function is shifted by a phase factor
$\exp(i\ch)$,
the off-diagonal elements of $\La$ shift by equal and opposite phases
$\exp(\pm 2 i \ch)$.

\section{Formalism}

For applications to the heavy-meson systems,
where less is known about CPT and T violation than in the $K$ system,
it is desirable to adopt a general 
parametrization of the effective hamiltonian $\La$ that is 
independent of phase conventions,\cite{ll}
valid for arbitrary size CPT and T violation,
model independent,
and expressed in terms of mass and decay rates insofar as possible.
An analysis shows that a practical parametrization 
permitting the clean representation of CPT- and T-violating quantities
can be obtained by expressing the two diagonal elements of $\La$
as the sum and difference of two complex numbers,
and the two off-diagonal elements
as the product and ratio of two other complex 
numbers.\cite{ak3}
A general expression for $\La$ can therefore be taken as:
\beq
\La = 
\half \De\la
\left( \begin{array}{lr}
U + \xi 
& 
\quad VW^{-1} 
\\ & \\
VW \quad 
& 
U - \xi 
\end{array}
\right),
\label{uvwx}
\eeq
where $UVW\xi$ are complex numbers that are dimensionless
by virtue of the prefactor $\De\la$.
Imposing that the trace of $\La$ is tr$~\La = \la$ 
and that its determinant is $\det \La = \la_a \la_b$
fixes the complex parameters $U$ and $V$:
\beq
U \equiv \la/\De\la, \quad 
V \equiv \sqrt{1 - \xi^2}.
\label{uvdef}
\eeq

The CPT and T properties of the effective hamiltonian \rf{uvwx} 
are contained in the complex numbers 
$W = w \exp (i\om)$,
$\xi = \Re\xi + i \Im \xi$.
Of the four real components,
the argument $\om$ of $W$ is physically irrelevant 
and can be freely dialed by the wave-function phase shifts described above.
The remaining three components are physical,
with $\Re\xi$ and $\Im\xi$ governing CPT violation
and the modulus $w\equiv |W|$ of $W$ governing T violation.
They are related to the components of $\La$ by
\beq
\xi = \De\La/\De\la,
\quad
w = \sqrt{|\La_{21}/\La_{12}|}.
\label{wxiexpr}
\eeq
If CPT is preserved $\Re\xi=\Im\xi=0$,
while if T is preserved $w = 1$.

The eigenstates of $\La$,
which are the physical states
of definite masses and decay rates, 
can be written as
\beq
\ket{P_a} =
\cN_a (\ket{P^0} + A \ket{\overline{P^0}}) , 
\quad
\ket{P_b} =
\cN_b (\ket{P^0} + B \ket{\overline{P^0}}) ,
\label{statesdef}
\eeq
with
\beq
A = (1 - \xi) W/V, \quad
B = -(1 + \xi) W/V.
\eeq
If unit-normalized states are desired,
the normalizations $\cN_a$, $\cN_b$ in Eq.\ \rf{statesdef}
take the form 
\beq
\cN_a = \exp(i \et_a)/\sqrt{1 + |A|^2}, 
\quad
\cN_b = \exp(i \et_b)/\sqrt{1 + |B|^2}, 
\label{norms}
\eea
where $\et_a$, $\et_b$ are free phases
that play no role in what follows.
For the special case with no CPT or T violation
($\xi = 0$, $w = 1$),
the states $\ket{P_a}$, $\ket{P_b}$ are CP eigenstates.
If the choice of phase convention $\om = \et_a = \et_b = 0$
is imposed,
Eq.\ \rf{statesdef} reduces to the usual form,
$\ket{P_{a,b}} = (\ket{P^0} \pm \ket{\overline{P^0}})/\sqrt 2$.

As an aside,
note that the $w\xi$ formalism above can be related to
other formalisms used in the literature
provided appropriate assumptions about the phase conventions
and the smallness of CP violation are 
made.\cite{ak3}
For instance,
in the $K$ system the widely 
adopted\cite{lw}
formalism involving $\ep_K$ and $\de_K$ 
is phase-convention dependent
and can be applied only if CPT and T violation are small.
Under the assumption of small violation
and in a special phase convention,
$\de_K$ is related to 
$\xi_K$ by $\xi_K \approx 2\de_K$.

For the heavy meson systems $D$, $B_d$, $B_s$,
the $w\xi$ formalism appears simpler to use than
other formalisms.
The three parameters for CP violation $w$, $\Re\xi$, $\Im\xi$
are dimensionless and independent of 
assumptions about the size of violations 
or about the choice of phases.
Since they are phenomenologically introduced,
they contain no model dependence.
However,
it is crucial to note that they need not be constant numbers.
In fact,
as outlined in the next section,
the assumption of constant $\xi$
often adopted for experimental and theoretical analyses 
represents a strong constraint on the generality of the formalism.
Moreover,
according to the CPT theorem,
the assumption of constant $\xi$ is inconsistent 
with the underlying basis of Lorentz-invariant quantum field theory.
If instead Lorentz violation is allowed within quantum field theory,
then $\xi$ is found to vary with the meson 4-momentum.
Although this may seem surprising at first sight,
in fact unconventional behavior for $\xi$
is to be expected because CPT violation is a fundamental effect.

\section{Theory}
\label{theory}

The standard-model 
extension\cite{kp,ck}
provides a general quantitative microscopic framework 
in the context of conventional quantum field theory
within which to study various effects of Lorentz and CPT violation.
As noted above,
many experiments with systems other than neutral mesons
have been performed to measure coefficients in this theory.
However,
to date none of these experiments has sensitivity
to the same sector of the standard-model extension
as neutral-meson oscillations,
basically because only the latter involve flavor 
changes.\cite{ak}

The dominant CPT-violating contributions to $\La$
can be calculated perturbatively in 
the coefficients for CPT and Lorentz violation
that appear in the standard-model extension.
These contributions are expectation values of 
perturbative interactions in the 
hamiltonian for the theory,\cite{kp}
evaluated with unperturbed wave functions
$\ket{P^0}$, $\ket{\overline{P^0}}$ as usual.
The hermiticity of the perturbation hamiltonian 
guarantees real contributions.

To find an expression for the parameter $\xi$,
one needs to derive the difference $\De\La =\La_{11} - \La_{22}$
of the diagonal terms of $\La$.
A calculation 
yields\cite{ak} 
\beq
\La \approx \be^\mu \De a_\mu ,
\label{dem}
\eeq
where $\be^\mu = \ga (1, \vec \be )$ is the four-velocity
of the meson state in the observer frame.
In this equation,
$\De a_\mu = r_{q_1}a^{q_1}_\mu - r_{q_2}a^{q_2}_\mu$,
where $a^{q_1}_\mu$, $a^{q_2}_\mu$
are coefficients for CPT and Lorentz violation 
for the two valence quarks in the $P^0$ meson.
They have mass dimension one,
and they arise from lagrangian terms
of the form $- a^q_\mu \overline{q} \ga^\mu q$,
where $q$ specifies the quark flavor.
The quantities $r_{q_1}$, $r_{q_2}$
emerge from normalization and quark-binding 
effects.\cite{kp}

Among the consequences of Lorentz and CPT violation 
are the 4-velocity and hence 4-momentum dependence 
appearing in Eq.\ \rf{dem}.
These establish the failure of the standard assumption of 
constant parameter $\xi$ for CPT violation.
In particular, 
the appearance of the 4-velocity implies that 
CPT observables will typically vary with the magnitude 
and orientation of the meson momentum.
This can have major consequences for experimental analyses,
since the meson momentum spectrum and angular distribution
now contribute directly in determining the experimental CPT 
reach.\cite{ak,ak2,ak3}

A crucial effect of the 4-momentum dependence
is the appearance of sidereal variations in some CPT 
observables.\cite{ak,ak2,ak3}
The point is that the vector $\De\vec a$ is constant,
while the Earth rotates in a celestial equatorial frame.
Since a laboratory frame is adopted for
the derivation of Eq.\ \rf{dem},
and since this frame is rotating,
observables can exhibit sidereal variations.
To display explicitly this sidereal-time dependence,
one can convert the expression \rf{dem} for $\De\La$
from the laboratory frame to a nonrotating frame.
Denote 
the spatial basis in the laboratory frame by 
$(\x,\y,\z)$
and that in the nonrotating frame by 
$(\X,\Y,\Z)$.
Choose the $\z$ axis in the laboratory frame 
for maximal convenience:
for example,
the beam direction is a natural option 
for the case of collimated mesons,
while the collision axis could be adopted in a collider.
Define the nonrotating-frame basis $(\X,\Y,\Z)$ 
to be consistent with celestial equatorial 
coordinates,\cite{ccexpt}
with $\Z$ aligned along the Earth's rotation axis.
Assume
$\cos{\ch}=\z\cdot\Z$ is nonzero,
as required for the observation of sidereal variations.
It follows that $\z$ precesses about $\Z$ with 
the Earth's sidereal frequency $\Om$.
The complete transformation between the two bases is in the
literature.\cite{ccexpt}
In particular,
any coefficient $\vec a$ for Lorentz violation
with laboratory-frame components $(a^1, a^2, a^3)$
has associated nonrotating-frame components $(a^X, a^Y, a^Z)$.
This transformation determines the sidereal variation of $\De \vec a$
and hence of $\De\La$.
The entire momentum and sidereal-time dependence 
of the parameter $\xi$ for CPT violation
in any $P$ system can then be extracted.

To express the final answer for $\xi$,
define $\th$ and $\ph$ to be standard polar coordinates
about the $\z$ axis in the laboratory frame.
These angles reduce to the usual detector polar coordinates
if the $\z$ axis is chosen along the detector axis.
In general,
the laboratory-frame 3-velocity of a $P$ meson can be written as
$\vec\be = \be (\sin\th\cos\ph, \sin\th\sin\ph, \cos\th)$.
The magnitude of the momentum 
is given by $p \equiv |\vec p| =\be m_P \ga(p)$,
where $\ga(p) = \sqrt{1 + p^2/m_P^2}$ as usual.
In terms of these quantities
and the sidereal time $\hat t$,
the result for $\xi$ 
becomes\cite{ak3}
\bea
\xi &\equiv &
\xi(\hat t, \vec p) \equiv \xi(\hat t, p, \th, \ph) 
\nonumber\\
&=& 
\fr 
{\ga( p)}
{\De \la} 
\bigl\{
\De a_0 
+ \be \De a_Z 
(\cos\th\cos\ch - \sin\th \cos\ph\sin\ch)
\nonumber\\
&&
\qquad
+\be \bigl[
\De a_Y (\cos\th\sin\ch 
+ \sin\th\cos\ph\cos\ch )
\nonumber\\
&&
\qquad \qquad
-\De a_X \sin\th\sin\ph 
\bigr] \sin\Om \hat t
\nonumber\\
&&
\qquad
+\be \bigl[
\De a_X (\cos\th\sin\ch 
+ \sin\th\cos\ph\cos\ch )
\nonumber\\
&&
\qquad\qquad
+\De a_Y \sin\th\sin\ph 
\bigr] \cos\Om \hat t
\bigr\} .
\label{xipt}
\eea

\section{Experimental Tests}
\label{expt}

The experimental challenge is to measure 
the four independent coefficients $\De a_\mu$ 
for CPT violation allowed by quantum field theory.
The result \rf{xipt} shows that suitable binning of data
in sidereal time, momentum magnitude, and orientation
has the potential to extract four independent bounds
from any observable that depends on $\xi$.
Note that each neutral-meson system can have different values of 
these coefficients.
Since the physics of each system is distinct 
by virtue of the distinct masses and decay rates,
a complete experimental analysis of CPT violation
requires four independent measurements in each system.

Consider the special case of semileptonic decays
into a final state $f$ or its conjugate state $\overline{f}$.
For simplicity,
disregard any violations of the 
$\De Q = \De S$,
$\De Q = \De C$,
or $\De Q = \De B$ rules.
Then,
the basic transition amplitudes can be taken as 
$\bra{f}T\ket{P^0} = F$,
$\bra {\overline f}T\ket {\overline{P^0}} = \overline F$,
$\bra{f}T\ket{\overline{P^0}} 
=\bra {\overline f}T\ket{P^0} = 0$.
The standard procedure can be applied to obtain the
various time-dependent decay amplitudes and probabilities.
Since the meson decays quickly relative to the Earth's sidereal period,
the dependence of $\xi$ on the meson proper time $t$ 
can be neglected.
The decay probabilities depend on the proper time,
as usual,
but in the presence of CPT violation
they also acquire sidereal time and momentum dependences 
from those of $\xi(\hat t, \vec p)$.

To illustrate the resulting effects for the case of uncorrelated mesons,
suppose direct CPT violation is negligible, 
so that $F^* = \ol F$.
An appropriate asymmetry sensitive to CPT violation is then
\bea
\cA^{\rm CPT}(t,\hat t,\vec p) &\equiv& 
\fr{
\overline{P}_{\overline{f}}(t,\hat t,\vec p) - P_f(t,\hat t,\vec p) 
}{
\overline{P}_{\overline{f}}(t,\hat t,\vec p) + P_f(t,\hat t,\vec p) 
}
\nonumber\\
&&
=\fr{
 2 \Re \xi \sinh\De\ga t/2
+ 2 \Im \xi \sin \De m t
}{
(1 + |\xi|^2) \cosh\De\ga t/2
+(1 - |\xi|^2) \cos\De m t
}.
\label{corrasymm}
\eea
This is understood to depend on $\hat t$, $\vec p$ 
through $\xi (\hat t, \vec p)$.
Independent measurements of the four coefficients $\De a_\mu$
can be obtained 
by various suitable averagings over 
$t$, $\hat t$, $p$, $\th$, $\ph$,
either before or after constructing the asymmetry \rf{corrasymm}.
For example,
if data are binned in $\hat t$
then it follows from Eq.\ \rf{xipt}
that measurements of the CPT coefficients $\De a_X$ and $\De a_Y$
are possible.
As another example,
binning in $\th$ separates the spatial and timelike components 
of $\De a_\mu$.

To date,
these ideas have been applied 
in experiments with the $K$ and $D$ systems.
For the $K$ system,
two independent CPT measurements of 
different combinations of the coefficients $\De a_\mu$
have been 
obtained,\cite{kexpt,ak}
one about $10^{-20}$ GeV on a linear combination of $\De a_0$ and $\De a_Z$,
and the other about $10^{-21}$ GeV on 
a combination of $\De a_X$ and $\De a_Y$.
The experiments in question
were performed with mesons highly collimated in the laboratory frame.
In this situation,
$\xi$ simplifies because
the 3-velocity takes the form $\vec\be = (0,0,\be )$.
Binning in $\hat t$ provides sensitivity to 
the equatorial components $\De a_X$, $\De a_Y$,
while averaging over $\hat t$ eliminates them altogether.
For the $D$ system,
preliminary sensitivity results 
for two independent bounds have also been obtained by the FOCUS 
experiment.\cite{dexpt}
Note that CPT bounds in the $D$ system are unique
in that the valence quarks involved are the $u$ and the $c$,
whereas the other neutral mesons involve the $d$, $s$, and $b$.

A different illustration is provided by the case of
correlated meson pairs produced by quarkonium decay 
into $f\overline{f}$.
The double-decay probability 
is a function of the proper decay times $t_1$, $t_2$,
the momenta $\vec p_1$, $\vec p_2$,
and the sidereal time $\hat t$.
The CPT properties of the two mesons in each decay 
typically are distinct
because the corresponding parameters $\xi_1$ and $\xi_2$ differ.
Since the time sum $t = t_1 + t_2$ is typically unobservable in practice,
an integration over $t$ is appropriate in deriving the
relevant probability $\Ga_{f\ol f}$.
It is then natual to define
a CPT-sensitive asymmetry $\cA^{\rm CPT}_{f\ol f}$
as a function of 
the difference $\De t = t_1 - t_2$ and the sum $\xi_1 + \xi_2$:
\bea
\cA^{\rm CPT}_{f\ol f}(\De t, \hat t, \vec p_1, \vec p_2)
&=&
\fr{
\Ga _{f\ol f}(\De t, \hat t, \vec p_1, \vec p_2) 
- \Ga _{f\ol f}(-\De t, \hat t, \vec p_1, \vec p_2)
}{
\Ga _{f\ol f}(\De t, \hat t, \vec p_1, \vec p_2)
+ \Ga _{f\ol f}(-\De t, \hat t, \vec p_1, \vec p_2)
}
\nonumber\\
&=&
\fr{
-\Re (\xi_1 + \xi_2)
\sinh \half \De\ga \De t
-\Im(\xi_1 + \xi_2)
\sin\De m\De t
}{
\cosh \half \De\ga \De t
+
\cos \De m\De t
}.
\nonumber\\
\label{sumxiasymm}
\eea
As in the previous asymmetry,
$\xi_1$, $\xi_2$ are understood to have
sidereal-time and momenta dependences,
so the attainable CPT reach can depend on the specific experiment.
Suppose,
for example,
that the quarkonium is created at rest in a symmetric collider.
The sum $\xi_1 + \xi_2 = 2 \ga (p)\De a_0/\De\la$
is then independent of $\De \vec a$,
and direct fitting of the data binned in $\De t$
allows a measurement of $\De a_0$.
If instead
the quarkonium is created in an asymmetric collider,
then $\xi_1 + \xi_2$ could be sensitive 
to all four coefficients $\De a_\mu$ for that neutral-meson system.
This implies that appropriate data binning
would allow up to four independent CPT measurements.
The existing asymmetric $B_d$ factories BaBar and BELLE
can undertake measurements of these types.

\section*{Acknowledgments}
This work was supported in part
by DOE grant DE-FG02-91ER40661.

\end{document}